\documentclass[conference]{IEEEtran}

\usepackage[pdftex]{graphicx}
\usepackage{amsmath}
\usepackage{eqparbox}

\begin{document}

\title{\LARGE Analysis of Fault Tolerant Multi-stage Switch Architecture for TSN}

\author{
	\IEEEauthorblockN{
			Adnan Ghaderi, Rahul Nandkumar Gore 	}

	\\
	\IEEEauthorblockA{
		Mälardalen University, Västerås, Sweden 
		\\
		 {\{adnan.ghaderi, rahul.nandkumar.gore\}}@mdu.se
	}
	
	}

\maketitle


\begin{abstract}
We conducted the feasibility analysis of utilizing a highly available multi-stage architecture for TSN switches used for sending high priority, mission-critical traffic within a bounded latency instead of traditional single-stage architectures. To verify the TSN functionality, we implemented the 'strict priority' feature. We evaluated the performance of both architectures on multiple parameters such as fault tolerance, packet latency, throughput, reliability, path length effectiveness, and cost per unit. The fault tolerance analysis demonstrated that the multi-stage architecture fairs better than the single-stage counterpart. The average latency and throughput performance of multi-stage architectures, although low, can be considered comparable with single-stage counterparts. However, the multi-stage architecture fails to meet the performance of single-stage architectures on parameters such as reliability, path length effectiveness, and cost-effectiveness. The improved fault tolerance comes at the cost of increased hardware resources, cost, and complexity. However, with the advent of cost-effective technologies in hardware design and efficient architecture designs, the multi-stage switching architecture-based TSN switches can be made reasonably comparable to single-stage switching TSN switches. This work gives initial confidence that the multi-stage architecture can be pursued further for safety-critical systems that require determinism and reliability in the communication of critical messages. 
\end{abstract}



\section{Introduction}
\begin{figure}[!t]
\centering
\includegraphics[width=3.5 in]{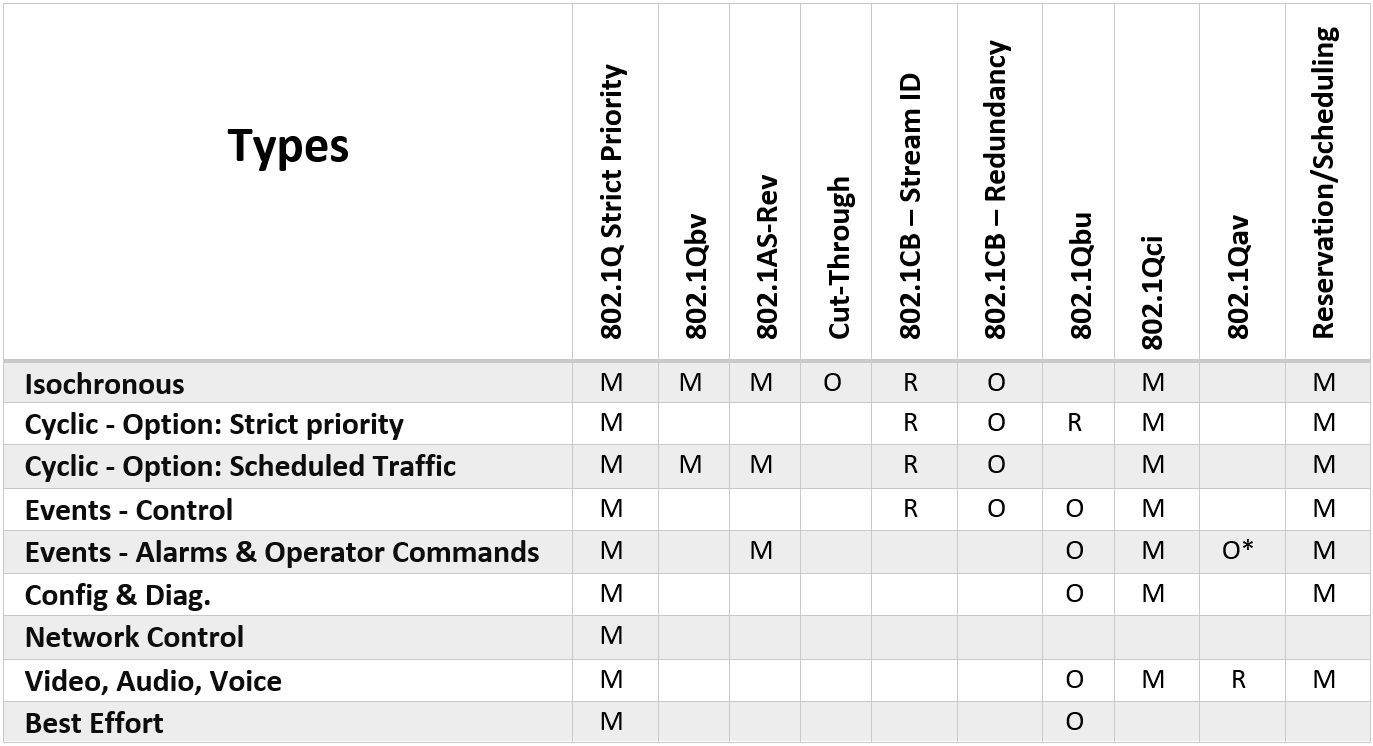}
\caption{Traffic types and TSN mechanisms (M = Mandatory,
O = Optional,
C = Conditional,
R = Recommended,
* = End devices)}
\label{fig: TSNtypes1}
\end{figure}

Owing to digitalization trends driven by advances in industry \texttt{4.0} and industrial IoT (Internet of Things), there is a growing need for industrial networks to support different types of traffic with varying priorities on the same network~\cite{7883994}. The prominent use case of such a scenario is to carry high-priority control traffic from process control and low-priority condition monitoring data to the cloud for predictive analysis on the same network. Future communication networks are envisioned to adopt TSN for the time-sensitive transmission of data over the deterministic Ethernet network~\cite{8695835}. 
TSN allows merging higher priority time-critical and non-prioritized traffic into a single network. 
Fig.~\ref{fig: TSNtypes1} shows the different types of traffic in industrial networks and TSN mechanisms to fulfill their QoS requirement~\cite{FlexManufact}.

\begin{figure}[!t]
\centering
\includegraphics[width=3.5 in]{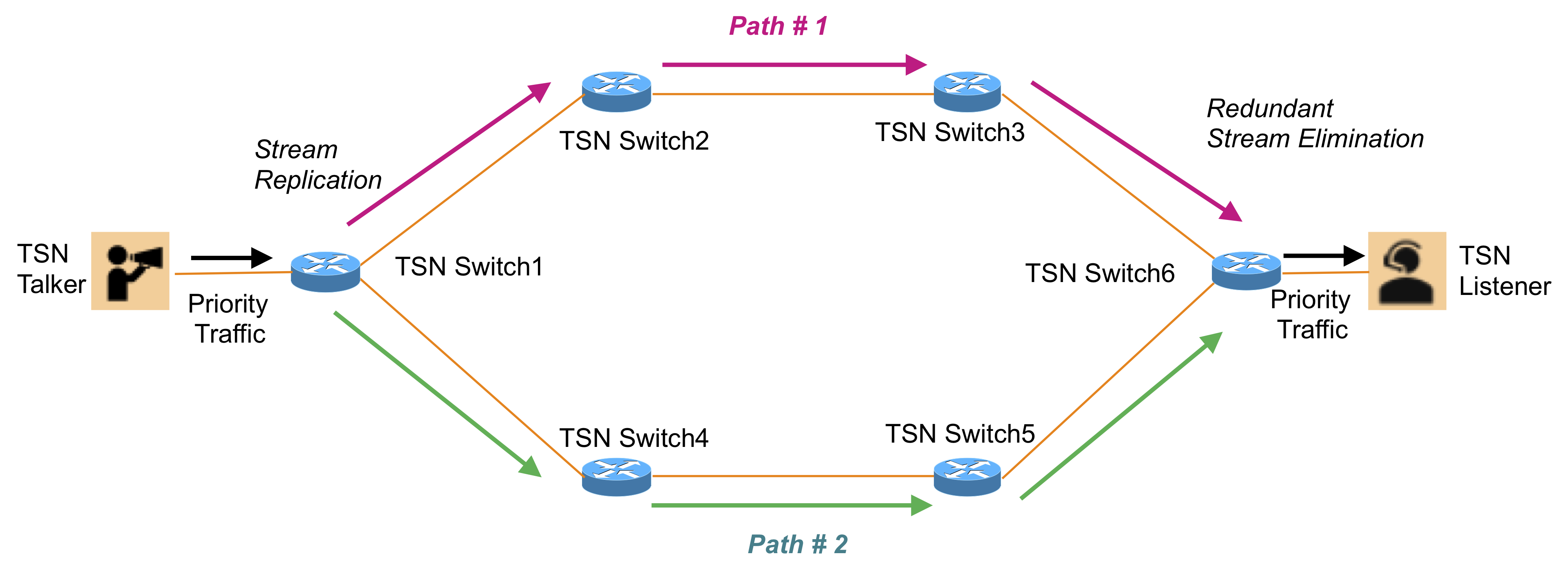}
\caption{Frame Replication and Elimination for Reliability (FRER) in IEEE 8021CB}
\label{fig: 8021CB}
\end{figure}

Safety-critical and safety-aware applications such as
control systems in the automotive industry and protection
mechanisms in electric substations warrant highly available
systems. Such systems require no frame loss and zero
recovery time delay. The redundancy or fault tolerance can be
achieved at the device, network switch, and communication link
level. One of the TSN standards IEEE 802.1CB provides redundancy at the communication stream level by tagging,
replicating and eliminating frames, as shown in Fig.~\ref{fig: 8021CB}. TSN switch 1 replicates the prioritized time-sensitive traffic from TSN talker device into two streams and sends them via two paths. TSN switch eliminates the redundant stream and sends only one stream to the receiver device. Thus it takes care of communication link failure by providing multiple and parallel network paths.

However, there is a single point failure in this system if the switch 1 or 6 stops functioning~\cite{8615374} as it directly affects the frame replication and elimination function. The most prominent failure mode of single-stage switches is the internal packet traverse-related failures.


If a redundancy can be provided at the TSN switch, it
would add immensely to overall system availability. The
safety-critical systems need such investigations to be carried
out for improved system availability. The redundancy
through fault tolerance is a common practice, but such
improvement can increase the effort, cost, or size of the system.

Citing the need to improve the overall availability of safety critical systems , this paper investigates the feasibility of using multi-stage architectures for TSN switch and comes up with a performance analysis of single and multi-stage architectures on various parameters. We present the analysis outcome with an intention to provide guidance on bigger design decisions such that whether to go for multi-stage architecture for TSN switch and at what cost.



The contribution of the paper is as follows: \\
(1) Availability analysis of TSN switch architecture for priority messages is a less explored area by researchers. In addition, assessing the possibility of using multi-stage architecture for TSN switches is not investigated so far. We investigated the less explored area and proposed a new multi-stage architecture for TSN switches that is suitable for safety-critical applications. \\
(2) We analyzed the performance of traditional single-stage and proposed multi-stage switching architectures using multiple parameters such as fault tolerance, reliability, path length, and cost. The analysis provides important design trade-offs required to make decisions on switch architecture.  

The paper is organized as follows: \\
Section II describes the single-stage and multi-stage switching architectures. Section III describes the multi-stage architecture based TSN switch. Section IV provides the performance analysis for single-stage and multi-stage switching architectures. Section V concludes the work and provides the future steps.

\section{Switch Architecture}
\subsection{Introduction} 

TSN brings the system intelligence into network devices, so the TSN switch is an integral part of the TSN ecosystem. TSN switch architecture needs to account for scalability, efficient memory management, higher throughput, better resource utilization, real-time performance, and redundancy~\cite{TSNswitch}. 

Among various options, shared bus, shared memory,
and the crossbar is the primary switching architectures used for different industrial applications. Initially, shared memory
and shared buses were used extensively. However, with high
performance and high-speed applications, a crossbar switch
grabbed more attention. Crossbar-based switches guarantee
non-blocking switching by connecting different input ports
to output ports simultaneously.

The crossbar switch's performance was hampered when
used for a bigger network with a higher number of ports.
With multiple input ports wanting to send data to the same
output ports, many packets would drop. Secondly, when
a cross-point of one input-output port pair connection is
failed, no data can be sent from input to output. The first
problem was resolved by employing buffers at input/output
ports or both. The second problem can be resolved by using
a multi-stage switch instead of a single stage. With multi-stage, if a cross-point is failed, there are other alternatives paths.

\subsection{Single-stage Switch Architecture}
\begin{figure}[t]
\centering
\includegraphics[width=3.5 in]{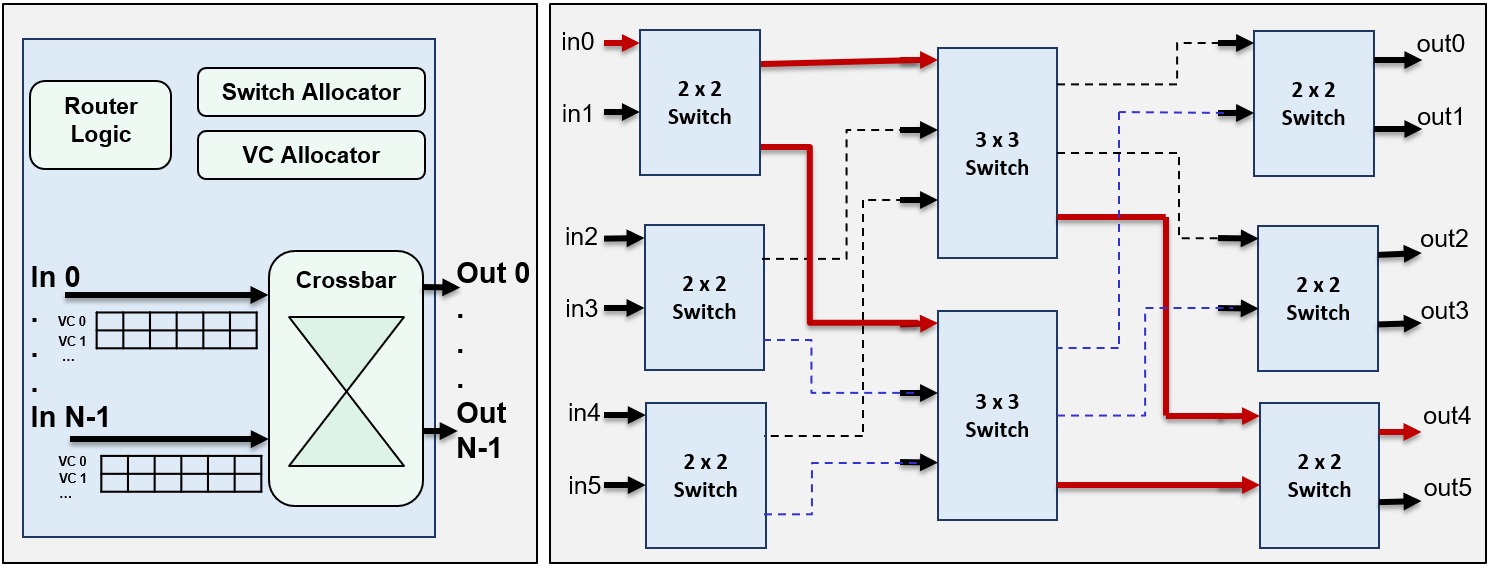}
\caption{(a) Single-stage (b) Multi-stage Switch Architecture}
\label{fig: SM}
\end{figure}

A Single-stage switch is composed of registers, switches, function units, and control logic that
collectively implement the routing and flow control functions required to buffer and
forward packets en route to their destinations. Although many router organizations exist,
Fig.~\ref{fig: SM}(a) shows a typical single-stage virtual-channel router.

Modern single-stage switches are pipelined at the packet level. Head packets proceed through pipeline
stages that perform RC (route computing) and VC (virtual channel) allocation, and all packets pass through SA (switch allocation) and ST (switch traversal) stages. Pipeline stalls occur if multiple packets compete for VC and SA allocation. Various arbitration schemes are used to deal with such scenarios and amicably allocate resources. 

For advancing a packet in a switch, RC is first to be performed to determine the output port (or ports) to which the packet can be forwarded. The packet then requests an output VC from the VC allocator. Once a route has been determined, and a virtual channel allocated, each packet is forwarded over this VC by allocating a time slot on the switch and output channel using the SA and forwarding the packet to the appropriate output unit during this time slot. Finally, the output unit forwards
the packet to the next router.

\subsection{Multi-stage Switch Architecture}
The multi-stage switch addresses the limitations of the
single-stage architecture to scale to a bigger network having
a large number of port requirements. A multi-stage architecture
includes one input stage, one output stage, and several
intermediate stages in the middle. Each stage comprises several
sub-switches, as shown in Fig.~\ref{fig: SM}(b). 
The advantage of this architecture is that one can design redundant paths between input and output port pairs. Interconnecting multi-stage networks are an attractive alternative to dedicated wiring because they allow scarce wiring resources to be shared by several low-duty-factor signals~\cite{8861540}.
\section{Multi-stage Architecture based TSN switch}
\begin{figure}[t]
\centering
\includegraphics[width=3.5 in,height = 2.1 in]{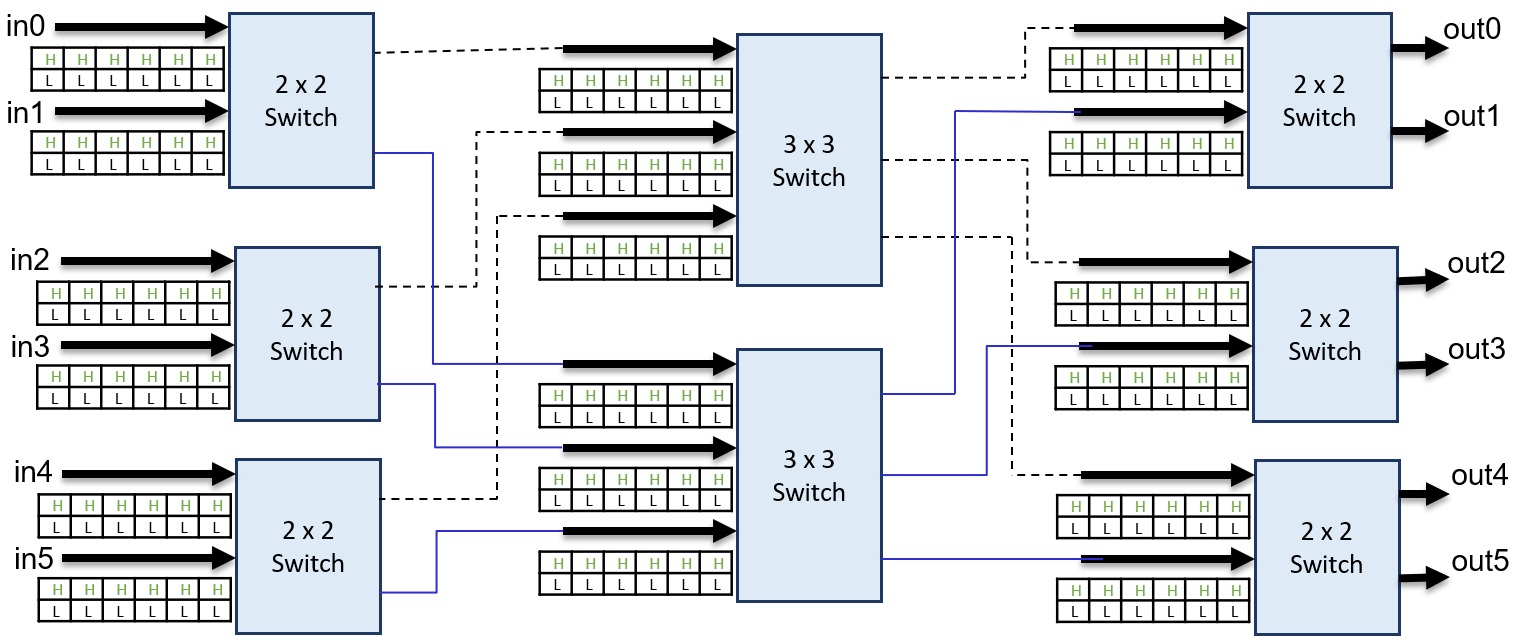}
\caption{Multi-stage switch architecture for priority based on VC}
\label{fig: Prio12}
\end{figure}

There are different approaches to implement multi-stage  architectures~\cite{dally}. Clos-network is a prominent multi-stage architecture that utilizes the switching resources optimally~\cite{clos1953study}. We analyzed a multi-stage switch architecture based on a clos-network, as shown in Fig.~\ref{fig: Prio12} for priority messages in safety-critical systems. 

We implemented one of the TSN features, a 'strict priority' for traffic by customizing VC and switch allocation functions within a switch. In modified architecture, the requests to high-priority packets are granted first, and then any remaining resources are allocated to lower-priority packets. Thus high priority traffic always gets preference over lower priority traffic in all the stages and traverses to the output port first. We proposed two multi-stage architectures to implement strict priority feature.\newline
1) Priority based on packets: The already tagged priority packets at all the ports and VCs were identified and given preference in VC and switch allocation.\newline
2) Priority based on VC: All the packets reaching a particular port and VC queue were granted access to VC and switch allocation on a priority basis compared to packets at other ports and VC queues.

 The first approach has a worst-case delay of one packet. If a low priority packet is under processing, then the high priority packet has to wait for that packet to reach the output. For this reason, we implemented a multi-stage architecture with priority based on VC as shown in Fig.~\ref{fig: Prio12}.
This architecture could be blocking, non-blocking.
It is rearrangeable based on the number of switches in each
stage and the number of ports for each sub-switches. The
proposed architecture can be scaled to any configuration
. However, with the increasing number of ports per switch or
number of switches in a particular stage, the routing and
flow control of packets becomes hugely challenging.
\section{Performance Evaluation}
Many parameters are applicable for comparing the performance of single and multi-stage architectures for TSN. Since worst-case latency and throughput are essential from TSN functionality point of view, we included them. Some of the critical performance parameters included are fault tolerance, reliability, path length effectiveness, and cost per unit. The examination of fault tolerance analysis provides deep observation of the possibility of using this architecture for higher availability. The TSN-specific parameters latency and throughput give confidence in terms of meeting message deadlines. We used BookSim simulator to implement the single and multi-stage TSN switch and perform the above analysis. The other parameters such as reliability, path effectiveness, and cost per unit help make trade-off decisions on whether to go for multi-stage architecture and cost. We used logical implementation architecture of single-stage and Benes-based multi-stage architecture for these theoretical analyses.  

We used a BookSim simulator to model the proposed multi-stage switch architecture and assess the performance in terms of latency, throughput, and fault tolerance. BookSim is a cycle-accurate simulator
for network-on-chip architectures. It features a modular design and offers a broad set of configurable network parameters in topology, routing algorithm, flow control, and router micro-architecture, including buffer management and allocation schemes. BookSim furthermore emphasizes detailed implementations of network components that accurately model the behavior of actual hardware~\cite{6557149}. We modeled the architectures shown in Fig.~\ref{fig: Prio12} and fed them with random uniform traffic. The routing and flow control functions were customized to implement the strict priority feature of TSN tools. The simulation results obtained from BookSim simulator are used to assess latency, throughput, and fault tolerance.

\begin{figure*}[t]
\centering
\includegraphics[width=7.0 in]{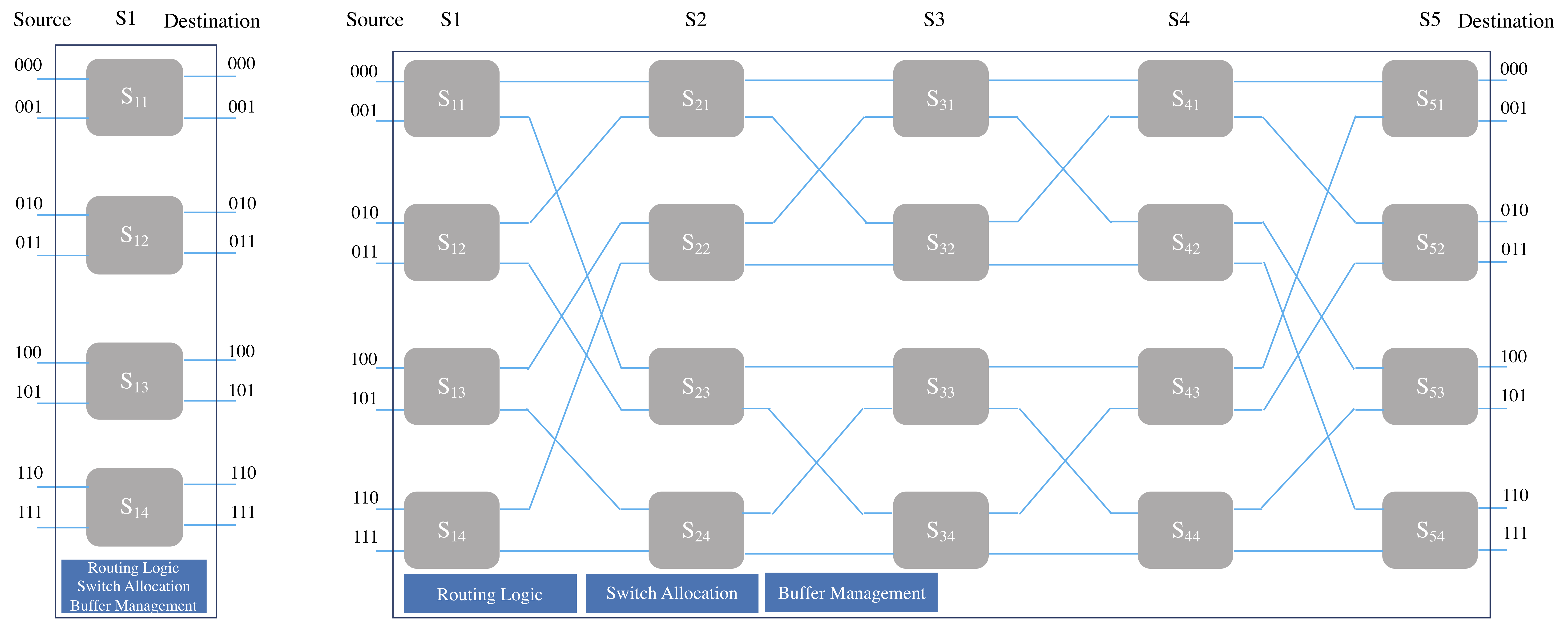}
\caption{Logical implementation of (a) Single-stage (b) Multi-stage Switch Architecture}
\label{fig: Benes}
\end{figure*}

In addition, we used logical implementation of single-stage and multi-stage architecture as shown in Fig.~\ref{fig: Benes}. Generally, a single-stage switch of size N×N (N inputs and N outputs) is built from switching elements of size 2 × 2, and the number of stages is equal to 1, and there are N/2 switching elements in the only stage. The network complexity is defined as the total number of switching elements of size 2 × 2. Therefore, the complexity of a single-stage is equal to O(N/2). However, a significant problem in the single-stage architecture is that there is only one path between each source–destination pair. Therefore, if one of the switching elements in the path fails, the network will be down.
As a result, multi-stage architectures such as Benes architecture~\cite{article789} were proposed. The multi-stage network has a topology that can be viewed as a connection of a baseline network and a reverse baseline network with the center stages overlapped. A multi-stage network of size 8×8 is shown in Fig.~\ref{fig: Benes}. The architecture of size N × N consists of $((2log_{2}N)$-$1)$ stages, and each stage is composed of N/2 switching elements of size 2×2. The network complexity of such an architecture is $N/2(2(log_{2}N)$-$1)$~\cite{article123}. The logical implementation of architectures was used to assess the performance in terms of reliability, path length effectiveness, and cost per unit. 

\subsection{Packet latency}
The time required to deliver a unit of data (usually a packet or message)
through the network, measured as the elapsed time between the injection
of the first bit at the source to the ejection of the last bit at the destination. 

\begin{figure}[t]
\centering
\includegraphics[width=3.5 in]{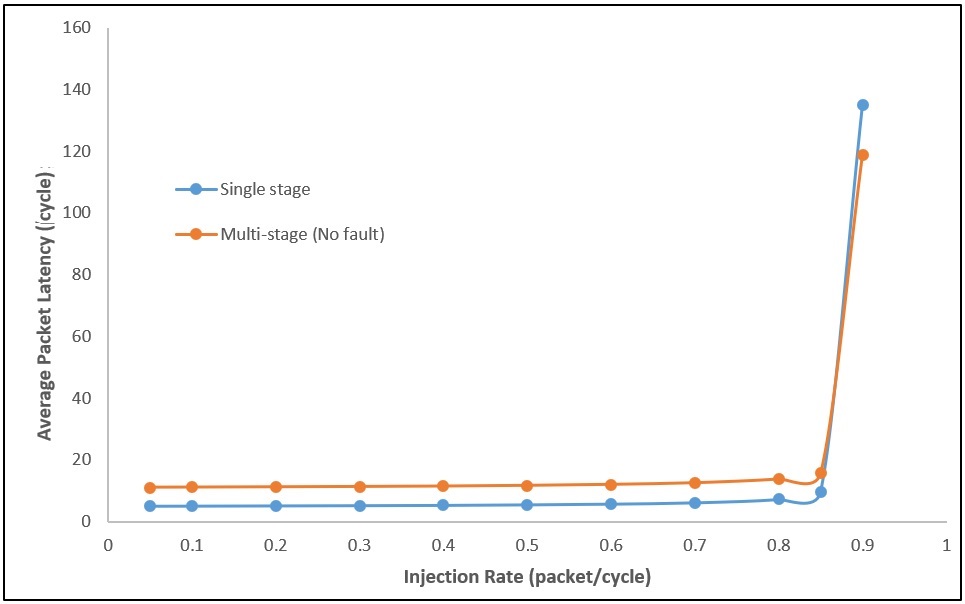}
\caption{Comparison of Average packet latency for single stage vs. multi stage architecture}
\label{fig: SvM1}
\end{figure}

\begin{figure}[t]
\centering
\includegraphics[width=3.5 in,height = 2.2 in]{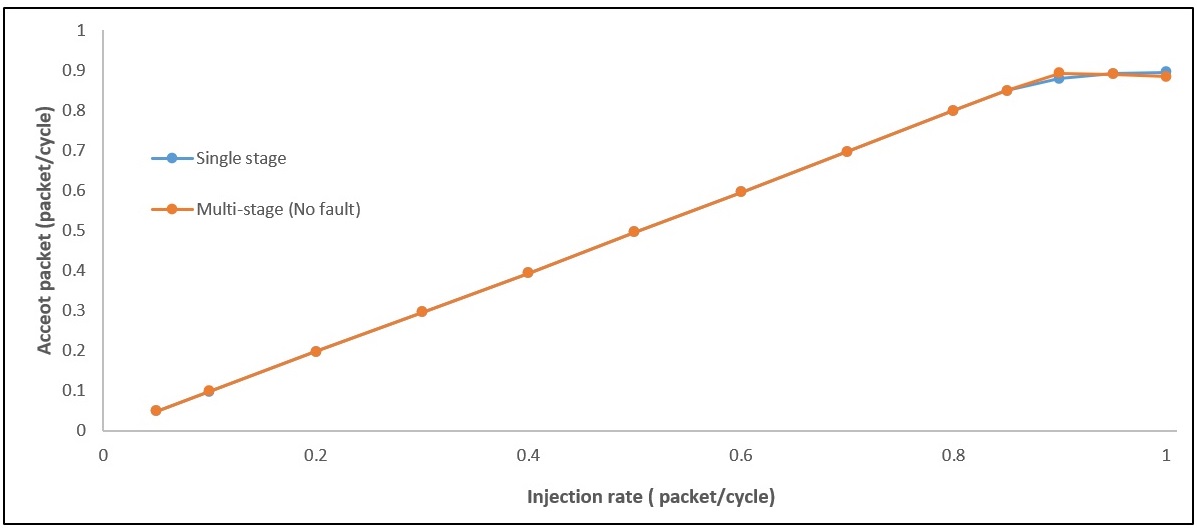}
\caption{Comparison of throughput for single stage and multi-stage architecture}
\label{fig: SvM2}
\end{figure}

Adding extra stages to single-stage switch architecture impacts packet latency more. This can be observed in Fig.~\ref{fig: SvM1} and Fig.~\ref{fig: SvM2}. The average packet latency for multi-stage is slightly more than for single-stage. Both the architectures exhibit the same worst-case performance. The worst-case latency in both architectures is around 0.85 packets/cycle.

\subsection{Throughput}
A resource is in saturation when the demands being placed on it
are beyond its capacity for servicing those demands. For example, a channel becomes saturated when the amount of data that wants to be routed over the channel (in bits/s) exceeds the channel's bandwidth. The saturation throughput of a network is the lowest rate of offered traffic for which some network resources are saturated.  

The throughput metric was obtained from the latency-throughput curves produced by simulations. The throughput figures are approximately the same for both architectures.

\subsection{Fault tolerance}
Fault tolerance is the ability of the network to perform
in the presence of one or more faults. For multi-stage
networks, operation in the presence of one or more faults is
an important attribute. Additionally, these
networks should degrade gracefully in the presence of faults. For
many systems, being single-point fault tolerance is sufficient
because if the probability of one fault is low, the chance of
simultaneous faults is extremely low~\cite{rel}.
A single-stage switch network cannot tolerate even a
single channel fault; hence, it was not simulated for this
analysis.
\begin{figure}[t]
\centering
\includegraphics[width=3.5 in]{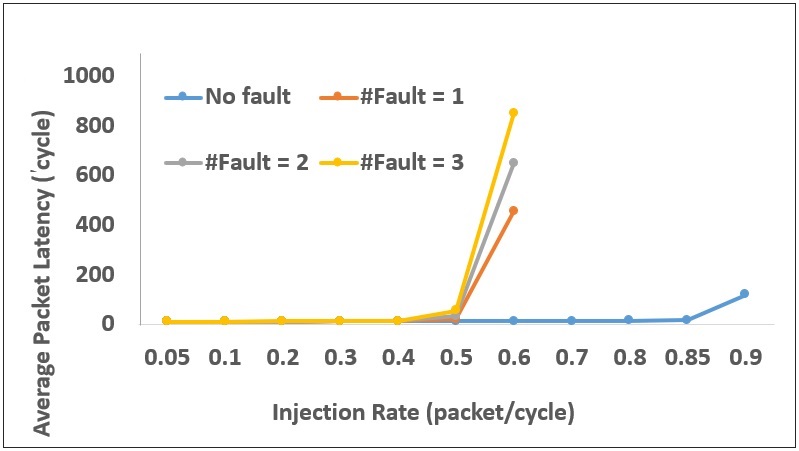}
\caption{Effect of faults on Average Packet Latency }
\label{fig: fault3a}
\end{figure}

\begin{figure}[t]
\centering
\includegraphics[width=3.5 in]{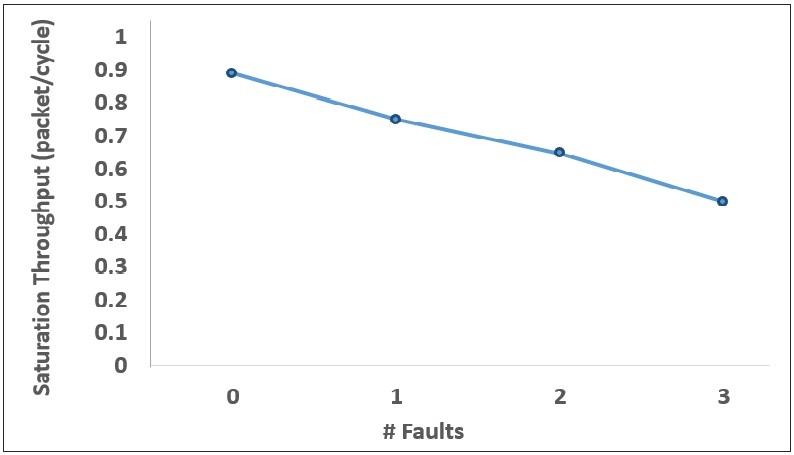}
\caption{Effect of faults on Saturation Throughput }
\label{fig: fault3b}
\end{figure}

In this experiment, a multi-stage network of Fig.~\ref{fig: Prio12} was simulated in BookSim with a variable number
of failed links forced through software. For each number of failures, the average packet latency and saturation throughput of the network under random uniform traffic was measured.

The average packet latency curve is shown in Fig.~\ref{fig: fault3a} indicates that architecture continues to function even after 3 simultaneous link failures. However, with an increasing number of faults, the saturation condition is reached earlier. 

The failed links also affect the saturation throughput, as shown in Fig.~\ref{fig: fault3b}. The throughput of the non-faulty network is just above 85\% of capacity, and the corresponding small drop in throughput illustrates the network's grace in the presence of a small number of faults. The network continues to remain resilient even as the number of faults grows to 3, with only a slight increase in the rate of
throughput degradation. Thus, we can say that a multi-stage
switch architecture degrades gracefully in the presence of
faults.

\subsection{Reliability}
Generally, reliability is the ability of a system to do and keep its functions in routine circumstances, as well as hostile or unexpected circumstances. Therefore, it is one of the essential requirements for an effective switch architecture.
Assuming that r is the probability of the 2x2 switching component being operational, system reliability of single-stage architecture in fig.~\ref{fig: Benes}(a) is given by

\begin{align} 
R_{ss}= 1-\sum_{i=1}^{N/2} (1 - r)
\end{align}

Most multi-stage architectures should be considered complex systems from the reliability point of view to determine the reliability~\cite{article456}. The reliability analysis of complex systems requires more complex calculations, so we use terminal reliability to calculate the reliability of multi-stage architecture in Fig.~\ref{fig: Benes}(b). 
Terminal reliability can be analyzed by considering a specific source-destination pair in the network. It is defined as the probability of successful communication between a source–destination pair. the terminal reliability of multi-stage architecture is given by

\begin{align} 
R_{ms}= r^{log_2N-1}(1-(1-(r^2(1-(1-r)^2)))^2
\end{align}

Suppose we use the typical figure of r=0.90 and check the reliability variation with increasing network size. In that case, we can see that the reliability performance of single-stage architecture is better than multi-stage architectures.

\subsection{Path Length Effectiveness}
Path length refers to the length of the communications path between the source and destination. Multiple paths of different path lengths are possible in a network. It can be measured by distance or by the number of intermediate switches. 

To take a closer look into paths between each source-destination pair, we have initially divided them into groups of basic paths. Then we have divided the basic paths into two groups: main paths and auxiliary paths. Basic paths (BPs) are those selected between any source and destination pair. Every request tries first the basic path; if it fails, then the same procedure is applied to other basic paths. Main paths (MP) are those used in usual conditions and have a shorter length than auxiliary paths (AP). Auxiliary paths, however, are those used when main paths are not available because they are busy or faulty. As mentioned earlier, their lengths are greater than the main ones. 

The total path length effectiveness (PLE) is calculated by,
\begin{align} 
Total\, PLE= \sum_{i=1}^{\infty}(\frac{NOMP_{n}}{LOMP_{n}}+\frac{NOAP_{n}}{LOAP_{n}}
\end{align}
where, NOMP and LOMP are number of main paths and length of main paths. NOAP and LOAP are number of auxiliary paths and length of auxiliary paths.
Using the formulas, it can be seen that the total PLE for single-stage architecture is 1 whereas it is 0.8 for multi-stage architecture.

\subsection{Cost per Unit}
Creating redundancy in the number of paths between each source–
destination pair is one of the main ways to improve the fault tolerance and reliability of switch architectures. However, redundancy usually imposes some costs on the network. Therefore, to create redundancy, a method will be suitable in case of reasonable costs. "Cost per unit" indicates how much we are spending per unit of performance~\cite{article123}. The following formula can calculate cost per unit:

\begin{align} 
Cost\, per\, unit = \frac{Total\, cost}{Total\, number\, of\, paths}
\end{align}

To estimate the cost of a network, one common method is to calculate the switching complexity with the assumption that the cost of a switch is proportional to the number of gates involved, which is roughly proportional to the number of cross-points within a switch. For example, a 2 × 2 switch has four units of hardware cost, whereas a 3 × 3 switch has nine units. \newline
The cost of single-stage network is calculated by the following equations:
\begin{align} 
Cost\, per\, unit = \frac{4*N/2}{N}\\
Cost\, per\, unit = 2
\end{align}

The cost of multi-stage network is calculated by the following equations:
\begin{align} 
Cost\, per\, unit = 2N(2(log_2N-1)-1)
\end{align}

Thus, it can be seen that the cost per unit increases with network size increase for multi-stage architectures, whereas it remains the same for single-stage architectures.
\begin{figure}[t]
\centering
\includegraphics[width=3.5 in]{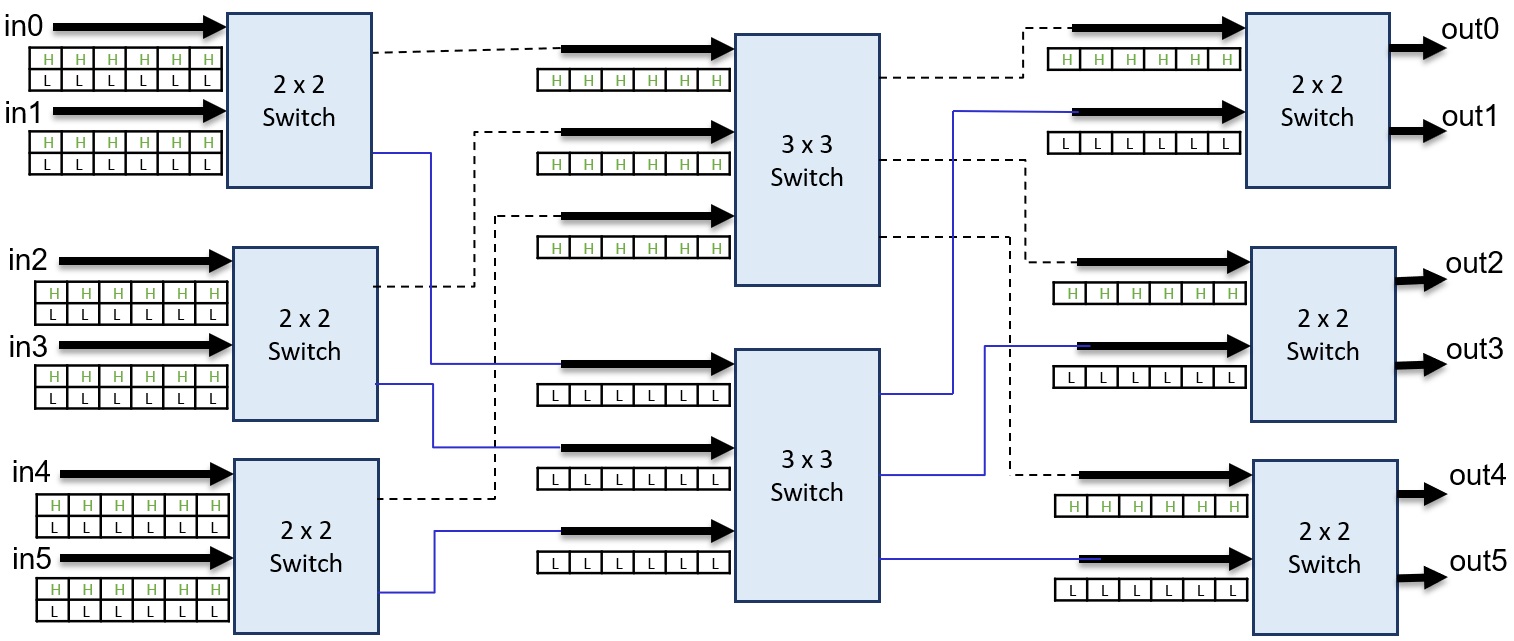}
\caption{Memory efficient multistage architecture }
\label{fig: Prio3}
\end{figure}

\subsection{A Memory efficient Scheme}
The proposed multi-stage switch architecture for the the configuration shown in Fig.~\ref{fig: Prio12} was modeled with design parameters VC per port = 2 and VC buffer size per port = 6. The total buffer memory requirement of the proposed architecture = N x 2 x 6 (stage 1) + N x 2 x 6 (stage 2) + N x 2 x 6 (stage 3) = 36N where N is number of ports in a particular stage.

\begin{figure}[t]
\centering
\includegraphics[width=3.5 in]{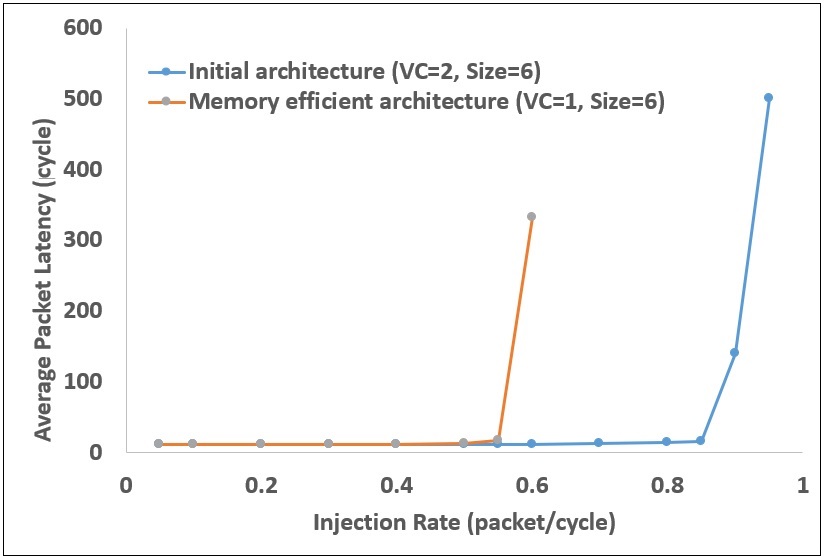}
\caption{Average Packet Latency of Memory Efficient architecture  }
\label{fig: mem3a}
\end{figure}

\begin{figure}[t]
\centering
\includegraphics[width=3.5 in]{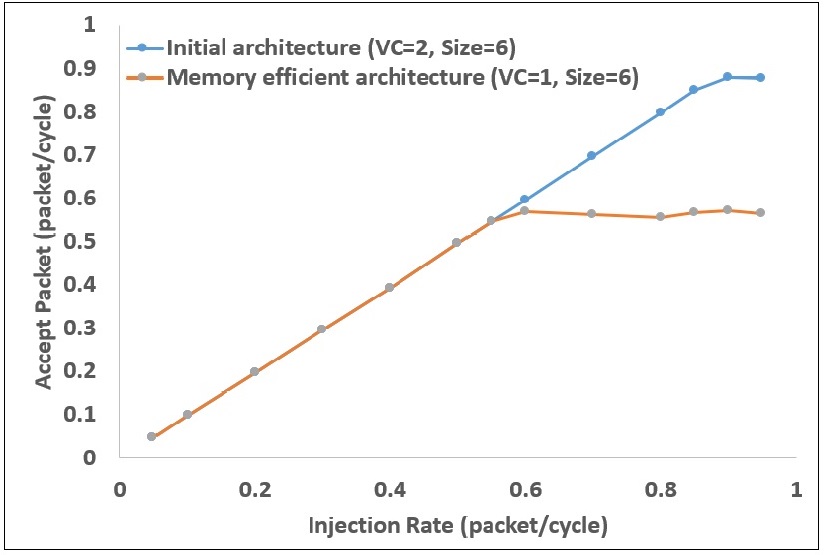}
\caption{Throughput of memory efficient architecture   }
\label{fig: mem3b}
\end{figure}

While the multi-stage switch architecture achieves fairly good fault tolerance performance, the buffer memory requirement is significantly high. The switch memories are expensive and hence are allocated for just optimal performance. With growing applications and network
devices, there is always a need to develop memory-efficient switch architectures. We developed a memory-efficient scheme, as shown in Fig.~\ref{fig: Prio3} for proposed multi-stage switch architecture. The scheme optimally rearranges the buffers allocated to the intermediate and last stages of the original architecture. The new modified switch architecture utilizes less number of buffers for communication of priority messages. 
The total buffer memory requirement of new architecture = N x 2 x 6 (stage 1) + N x 6 (stage 2) + N x 6 (stage 3) = 24N where N is the number of ports in a stage. Thus, we achieve around a 33\% reduction in buffer memory with new memory-efficient architecture.

We simulated both the multi-stage architectures in BookSim by feeding them high and low priority traffic simultaneously. By varying the injection rate of packets at input nodes, we obtained the average packet latency and throughput values for both architectures. 


The comparison analysis from Fig.~\ref{fig: mem3a} and Fig.~\ref{fig: mem3b} shows that the no-load packet latency for both the architectures is approximately the same but the memory-efficient architecture reaches the saturation quite earlier (0.55 packets/cycle injection rate) compared to initial architecture (0.85 packets/cycle injection rate). Thus the memory-efficient architecture reduces the buffer size requirement however at the cost of a lower saturation point. It is a decision for the switch designers to choose the right variant of architecture that meets a particular application requirement. 

\section{Conclusion and Future Work}
We conducted the feasibility analysis of utilizing a highly available multi-stage architecture for TSN switches  used for sending high priority traffic within a bounded latency instead of traditional single-stage architectures. We evaluated the performance of both architectures on multiple parameters such as fault tolerance, packet latency, throughput, reliability, path length effectiveness, and cost per unit. The fault tolerance analysis demonstrated that the multi-stage architecture fairs better than the single-stage counterpart. The average latency and throughput performance of multi-stage architectures, although low, can be considered comparable with single-stage counterparts. However, the multi-stage architecture fails to meet the performance of single-stage architectures on parameters such as reliability, path length effectiveness, and cost-effectiveness. The improved fault tolerance comes at the cost of increased hardware resources, cost, and complexity. However, with the advent of cost-effective technologies in hardware design and efficient architecture designs, the multi-stage architecture-based TSN switches can be made reasonably comparable to single-stage switching TSN switches.
To verify the TSN functionality, we implemented the
'strict priority' feature. This work gives confidence
that the proposed architecture can be pursued further
for safety-critical systems that require determinism and
reliability in the communication of critical messages. Further
TSN tool-sets such as path control, reservation, stream
filtering, and policing can be implemented in a future course.

\bibliographystyle{IEEEtran}
\bibliography{IEEEabrv,IMS2014_rectifier_short}

\begin{thebibliography}{10}
\providecommand{\url}[1]{#1}
\csname url@samestyle\endcsname
\providecommand{\newblock}{\relax}
\providecommand{\bibinfo}[2]{#2}
\providecommand{\BIBentrySTDinterwordspacing}{\spaceskip=0pt\relax}
\providecommand{\BIBentryALTinterwordstretchfactor}{4}
\providecommand{\BIBentryALTinterwordspacing}{\spaceskip=\fontdimen2\font plus
\BIBentryALTinterwordstretchfactor\fontdimen3\font minus
  \fontdimen4\font\relax}
\providecommand{\BIBforeignlanguage}[2]{{%
\expandafter\ifx\csname l@#1\endcsname\relax
\typeout{** WARNING: IEEEtran.bst: No hyphenation pattern has been}%
\typeout{** loaded for the language `#1'. Using the pattern for}%
\typeout{** the default language instead.}%
\else
\language=\csname l@#1\endcsname
\fi
#2}}
\providecommand{\BIBdecl}{\relax}
\BIBdecl

\bibitem{7883994}
M.~{Wollschlaeger}, T.~{Sauter}, and J.~{Jasperneite}, ``The future of
  industrial communication: Automation networks in the era of the internet of
  things and industry 4.0,'' \emph{IEEE Industrial Electronics Magazine},
  vol.~11, no.~1, pp. 17--27, 2017.

\bibitem{8695835}
L.~{Lo Bello} and W.~{Steiner}, ``A perspective on ieee time-sensitive
  networking for industrial communication and automation systems,''
  \emph{Proceedings of the IEEE}, vol. 107, no.~6, pp. 1094--1120, 2019.

\bibitem{FlexManufact}
P.~Didier and R.~Blair, ``{Time Sensitive Networks - Update from the IIC
  Testbed for Flexible Manufacturing},'' ODVA, Industry Conference, Tech. Rep.,
  October 2018.

\bibitem{8615374}
M.~{Pahlevan} and R.~{Obermaisser}, ``Redundancy management for safety-critical
  applications with time sensitive networking,'' in \emph{2018 28th
  International Telecommunication Networks and Applications Conference
  (ITNAC)}, 2018, pp. 1--7.

\bibitem{TSNswitch}
A.~Ghaderi, M.~Daneshtalab, M.~Ashjaei, M.~Loni, S.~Mubeen, and M.~Sjödin,
  ``Design challenges in hardware development of time-sensitive networking: A
  research plan,'' 09 2019.

\bibitem{8861540}
F.~{Montano}, T.~{Ould-Bachir}, J.~{Mahseredjian}, and J.~P. {David}, ``A
  low-latency reconfigurable multistage interconnection network,'' in
  \emph{2019 IEEE Canadian Conference of Electrical and Computer Engineering
  (CCECE)}, 2019, pp. 1--4.

\bibitem{dally}
\BIBentryALTinterwordspacing
W.~Dally, W.~James, B.~Towles, and B.~Patrick, \emph{Principles and Practices
  of Interconnection Networks}, ser. The Morgan Kaufmann Computer A.\hskip 1em
  plus 0.5em minus 0.4em\relax Elsevier Science, 2004. [Online]. Available:
  \url{https://books.google.se/books?id=oOqpcB5191sC}
\BIBentrySTDinterwordspacing

\bibitem{clos1953study}
C.~Clos, ``A study of non-blocking switching networks,'' \emph{Bell System
  Technical Journal}, vol.~32, no.~2, pp. 406--424, 1953.

\bibitem{6557149}
{Nan Jiang}, D.~U. {Becker}, G.~{Michelogiannakis}, J.~{Balfour}, B.~{Towles},
  D.~E. {Shaw}, J.~{Kim}, and W.~J. {Dally}, ``A detailed and flexible
  cycle-accurate network-on-chip simulator,'' in \emph{2013 IEEE International
  Symposium on Performance Analysis of Systems and Software (ISPASS)}, 2013,
  pp. 86--96.

\bibitem{article789}
K.~Aghakhani and A.~Karimi, ``A novel routing algorithm in benes networks,''
  \emph{International Journal of Educational Advancement}, vol.~7, p. 2016, 10
  2016.

\bibitem{article123}
F.~Bistouni and M.~Jahanshahi, ``Improved extra group network: A new
  fault-tolerant multistage interconnection network,'' \emph{The Journal of
  Supercomputing}, vol.~69, pp. 161--199, 07 2014.

\bibitem{rel}
N.~A.~M. Yunus, M.~Othman, Z.~M. Hanapi, and K.~Y. Lun, ``Reliability review of
  interconnection networks,'' \emph{IETE Technical Review}, vol.~33, no.~6, pp.
  596--606, 2016.

\bibitem{article456}
F.~Bistouni and M.~Jahanshahi, ``Impact of raising switching stages on the
  reliability of interconnection networks,'' \emph{Journal of the Institute of
  Electronics and Computer}, vol.~2, pp. 99--126, 01 2020.

\end{thebibliography}

\end{document}